\documentclass[aps,prl,superscriptaddress, twocolumn, showpacs]{revtex4-1}
\usepackage{graphicx}
\usepackage{dcolumn}
\usepackage{bm}
\usepackage{amsmath}

\usepackage{color}
\usepackage[flushleft]{threeparttable}
\usepackage{bm,graphicx,hyperref}
\usepackage{cancel}
\hypersetup{%
  breaklinks = {true},
  citecolor = {blue},
  colorlinks = {true},
  linkcolor = {red},
}
\usepackage{lineno}

\begin{document}

\title{Two-dimensional Square Buckled Rashba Lead Chalcogenides}

\author{Paul~Z.~Hanakata}
\affiliation{Department of Physics, Boston University, Boston, MA 
02215}
\email{hanakata@bu.edu}

\author{A.~S.~Rodin} \affiliation{Centre for Advanced 2D
  Materials and Graphene Research Centre, National University of
  Singapore, 6 Science Drive 2, 117546, Singapore}

\author{Alexandra~Carvalho} \affiliation{Centre for Advanced 2D
  Materials and Graphene Research Centre, National University of
  Singapore, 6 Science Drive 2, 117546, Singapore}

\author{Harold~S.~Park}
\affiliation{Department of Mechanical Engineering, Boston University, Boston, MA 
02215}

\author{David~K.~Campbell}
\email{dkcampbe@bu.edu}
\affiliation{Department of Physics, Boston University, Boston, MA  02215}

\author{A.~H.~Castro Neto} \affiliation{Centre for Advanced 2D
  Materials and Graphene Research Centre, National University of
  Singapore, 6 Science Drive 2, 117546, Singapore}

\date{3 October, 2017}
\begin{abstract}
  We propose the lead sulphide (PbS) monolayer as a 2D semiconductor
  with a large Rashba-like spin-orbit effect controlled by the
  out-of-plane buckling.  The buckled PbS conduction band is found to
  possess Rashba-like dispersion and spin texture at the $M$ and
  $\Gamma$ points, with large effective Rashba parameters of
  $\lambda\sim5$~eV\AA~and $\lambda\sim1$~eV\AA, respectively.  Using
  a tight-binding formalism, we show that the Rashba effect originates
  from the very large spin-orbit interaction and the hopping term that
  mixes the in-plane and out-of-plane $p$ orbitals of Pb and S
  atoms. The latter, which depends on the buckling angle, can be
  controlled by applying strain to vary the spin texture as well as
  the Rashba parameter at $\Gamma$ and $M$. Our density functional
  theory results together with tight-binding formalism provide a
  unifying framework for designing Rashba monolayers and for
  manipulating their spin properties.
\end{abstract}


\maketitle

\emph{Introduction--} Over the past two decades there has been a
growing interest in materials with strong spin-orbit interaction
(SOI), as they are of a profound importance for fundamental
understanding of quantum phenomena at the atomic level and
applications to spintronics.  This relativistic interaction is linked
to important effects such as Rashba, Zeeman, spin-Hall effect, and
topological insulator (TI) states~\cite{ishizaka-NatMat-10-521-2011,
  sakano-PRL-110-107204-2014, kim-PRL-115-086802-2015,
  liu-NanoLett-15-2657-2015}.

The spin-orbit splitting of the bands occurs in crystals without
inversion symmetry, where it is known as Dresselhaus effect and in 2D
structures or surfaces, where it is known as Rashba effect, even
though these can be seen as different manifestations of the same
phenomenon~\cite{zhang-NatPhys-10-387-2014}.  However, suitable
atomically thin 2D materials with a large Rashba coefficient are hard
to find.  To have Rashba-type spin splitting there are two key
properties that should present: strong SOI and broken inversion
symmetry.  In graphene and non-polar two-dimensional materials, such
as transition metal dichalcogenides, breaking inversion symmetry is
often achieved by application of out-of-plane electric fields or
through interfacial effects~\cite{hongki-PRB-74-165310-2006,
  yuan-NatPhys-9-563-2013, avsar-NatComm-5-4875-2013}. Unfortunately,
the respective spin splitting in graphene is rather small, rendering
the spin polarization unusable at room temperature.  Group IV and
III-V binary monolayers (e.g SiGe and GaAs) with buckled hexagonal
geometry were found to have a Rashba-like spin texture; the band
splitting, however, has a Zeeman-like
splitting~\cite{sante-PRB-91-161401-2015}. Spin-splitting in WSe$_2$
monolayer is also of Zeeman-type due to the out-of plane mirror
symmetry ($M_z:z\rightarrow-z$) suppressing the Rashba
term~\cite{yuan-NatPhys-9-563-2013}. Transition metal dichalcogenides
with asymmetric surfaces, e.g. WSeTe, have a sizable Rashba
splitting, but this does not coincide with the direct
bandgap~\cite{yao-PhysRevB-95-165401}.  A Rashba-type effect has been
measured in few-layer samples of the topological insulator
Bi$_2$Se$_3$, but this is attributed to the interactions with the
substrate~\cite{zhang-NatPhys-6-584-2010}.

Recently, we proposed that a Rashba-like splitting can also be
obtained in bucked heavy metal square lattices, where it is controlled
by out-of-plane buckling and/or electric dipole~\cite{rodin-PRB-96-115450-2017}.
However, materials in this class are almost always metals, which
reduces the ways in which spins can be manipulated.

In addition to study of spin splitting and texture in materials with
strong SOI, several works have also investigated the orbital switching
in topological insulators~\cite{cao-NatPhys-9-499-2013,
  zhang-PRL-111-066801-2013} and in {\it hexagonal} 3D Rashba
semiconductors~\cite{liu-PRB-94-125207-2016,
  bawden-scieAdv-1-e1500495-2015}. Specifically, Cao et al. found that
below the Dirac point the wavefunctions are more radial while above
the Dirac point the wavefunctions are more
tangential~\cite{cao-NatPhys-9-499-2013}. However, further studies for
materials with different geometry (e.g square) are still lacking.

Very recently, several studies have investigated topological
properties of the rock salt structure materials, such as PbX (X=Se, S,
Te), in both monolayer and bilayer forms with no
buckling~\cite{kim-PRL-115-086802-2015, liu-NanoLett-15-2657-2015,
  wan-AdvMat-29-1521-2017}. In particular, Chang et al. have
successfully grown few-layer SnTe and
PbTe~\cite{chang-science-353-274-2016}~\footnote{Chang et al. reported
  PbTe data in the supplementary section.}. In this article, we study
two-dimensional (2D) lead chalcogenide PbX (X=S, Se, Te) monolayers in
square geometry with two atoms per primitive cell.  For definiteness,
we focus on lead sulfide PbS, but similar effects can be found for
other lead chalcogenides and even heavy
metals~\cite{rodin-PRB-96-115450-2017}.

Using density functional theory (DFT), we find that buckled PbS
monolayer possesses a strong Rashba splitting. In this polar material,
the buckling direction can be reversed, leading to the reversal of the
spin texture. Based on our DFT results we develop a tight binding
formulation of the buckled and planar 2D square lattice for PbS which
is generally applicable for other similar materials ({\it {e.g}}, PbSe
and PbTe). With this formalism, we are able to understand how the
Rashba spitting depends on SOI strength, which in turn depends on the
atomic species and the buckling angle, similar to the case of heavy
metal square lattices~\cite{rodin-PRB-96-115450-2017}. Moreover, our
theory provides a new understanding of how spins and orbitals are
coupled and how they can be controlled. These all together provide
guidelines for designing and manipulating orbital-spin effects in
Rashba monolayers.

\emph{Methods--} Our findings are based on density functional theory
(DFT) calculations implemented in the {\sc Quantum ESPRESSO}
package~\cite{QE-2009}.  We employed Projector Augmented-Wave (PAW)
type pseudopotentials with Perdew-Burke-Ernzerhof (PBE) within the
generalized gradient approximation (GGA) for the exchange and
correlation functional~\cite{pbe-PRL-77-3865-1996}. The Kohn-Sham
orbitals were expanded in a plane-wave basis with a cutoff energy of
100 Ry, and for the charge density a cutoff of 200 Ry was used.  A
$k$-point grid sampling grid was generated using the Monkhorst-Pack
scheme with 16$\times$16$\times$1 points~\cite{monkhorstPRB1976}, and
a finer regular grid of 40$\times$40$\times$1 was used for spin
texture calculations. We used the modern theory of
polarization~\cite{vanderbilt-PRB-47-1651-1993} to calculate the
spontaneous polarization implemented in the {\sc Quantum ESPRESSO}
package~\cite{QE-2009}.  To compare the electric polarization of
monolayer PbS to the typical bulk ferroelectrics, we approximate the
thickness as twice the distance between S and Pb atom which is roughly
half of the lattice constant of bulk PbS. Similar approximations have
also been used in other several works~\cite{hanakata-PRB-94-035304-2016,
  fei-PRL-117-097601-2016, wang-2DMat-4-015042-2017}.

For electronic band structure calculations, the spin orbit interaction
was included using noncollinear calculations with fully relativistic
pseudopotentials. To apply biaxial strains, we varied the in-plane
lattice constants and let the system relax until the stress
perpendicular to the plane is less than 0.01~GPa.

\emph{Structure, bistability, and ferroelectricity--}
\begin{figure*}
\includegraphics[width=16cm]{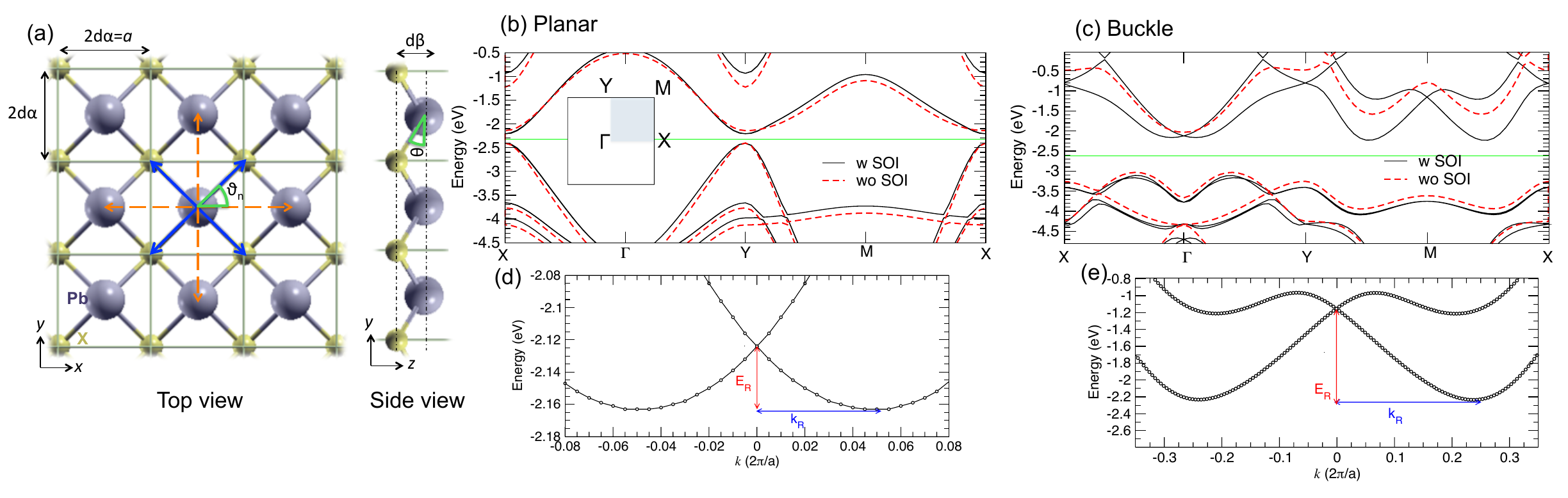}
\caption{(a) Structural visualizations of buckled PbS
  monolayer. Buckling angle $\theta=0$ ($\beta=0$) for planar
  structure. Blue and orange arrows indicate vectors connecting Pb and
  its first and second nearest neighbors, respectively. Band structure
  of monolayer PbS in planar (b) and buckled structure (c) along the
  high symmetry lines of Brillouin zone. Green lines indicate Fermi
  energy. There is no splitting in the planar structure because of
  inversion symmetry. In contrast, there is no mirror-plane in $z$ for
  buckled structure resulting broken inversion symmetry, and this
  leads to band-splitting. The calculated Rashba parameter at $\Gamma$
  ($M$) gives rise to a larger energy splitting between bands than
  other giant Rashba materials. Rashba-like dispersion at $\Gamma$ (d)
  and $M$ point (e).}
\label{fig:fig1}
\end{figure*}
Our first principles calculations show that PbX monolayer has a
buckled structure, which is a minimum of the energy surface, whereas
the planar structure is a saddle point of the energy
surface~\footnote{See Supplemental Material at [URL will be inserted
  by publisher] for the energy surface, projected density of states
  (PDOS), and full Hamiltonian\label{foot1}}. We found that the optimized buckled
structure of PbS has a lower enthalpy of 120~meV compared to that of
optimized planar structure. The lattice constant $a$ and buckling
angle $\theta$ for buckled (planar) structure are 3.74\AA~(4.01\AA)
and 21.6$^{\circ}$ (0$^{\circ}$), respectively. The optimized planar
lattice constant is close to the value reported in study of planar
PbS~\cite{wan-AdvMat-29-1521-2017}.

The energy barrier between the planar (paraelectric) and buckled
(ferroelectric) is obtained by displacing the Pb and S atoms in the
$z$ direction while keeping the lattice parameters fixed at the values
optimized for the the buckled (ferroelectric) phase. Using the fixed
ferroelectric (buckled) lattice parameters, the energy barrier is 764
meV and the spontaneous polarization is $Pol=0.2$~C/m$^2$. Since the
calculation is carried out keeping the lattice parameters fixed at the
values optimized for the buckled phase, the relative energy of the
paraelectric phase is overestimated. In fact, potential energy
barriers in ferroelectric materials are usually strain dependent. For
instance, Wang and Qian have shown that energy barriers in
ferroelectric SnS, SnSe, GeS, and GeSe monolayers may increase or
decrease depending on the strains~\cite{wang-2DMat-4-015042-2017}. To
support our argument, we also calculated the path where the optimized
paraelectric (planar) phase is used as the initial configuration. When
the lattice parameters are fixed at the optimized paraelectric phase,
the energy barrier is 51 meV and the spontaneous polarization is
$Pol=0.1$~C/m$^2$, as shown in the Supplemental Materia. By fitting
the energy surface to fourth order
polynomial~\cite{beckman-PRB-79-144124-2009}, we can calculate the
coercive field given by $E_c = (4/3)^{(3/2)} E_{\rm barrier}/Pol$.
The calculated coercive field with the starting configuration from
paraelectric (planar) and ferroelectric (buckled) are, $\sim 1$~V/nm
and $\sim 10$~V/nm respectively. Applied electric fields of
$\sim 1$~V/nm are achievable in current 2D
experiments~\cite{radisavljevic-NatNanotechnology-6-147-2011}. This
suggests that PbS is suitable for a ferroelectric device as long as it
is grown on its planar phase.
                
\emph{Band structure--}
Next we compare the band structure of planar PbS (PbS-$p$) and buckled
PbS (PbS-$b$). PbS-$p$ is a direct gap semiconductor with a small
bandgap of 0.2~eV. Because of the inversion symmetry, no
spin-splitting is observed. PbS-$b$ is an indirect-gap semiconductor
in which the minimum energy of the lowest conduction band is located
near the $M$-point and the maximum energy of the highest valence band
is located near the $\Gamma$-point.  At both the $M$ and $\Gamma$
points, the conduction band shows a sizable Rashba splitting.  The
effective Rashba parameters, given by $\lambda=2E_R/k_R$, where $E_R$
is the difference between the lowest energy of upper band and lower
band, $k_R$ is the shift in momentum space relative to the cone axis,
are $\lambda=1.03$~eV\AA~at $\Gamma$ (Fig.~\ref{fig:fig1}(d)) and
$\lambda=5.10$~eV\AA~at $M$ (Fig.~\ref{fig:fig1} (e)).  These values
are comparable to those of three-dimensional (3D) giant Rashba
materials~\cite{ishizaka-NatMat-10-521-2011,
  sakano-PRL-110-107204-2014, liebmann-AdvMat-28-560-2016}.

\emph{Origin of the spin splitting: a tight-binding formulation--}
Next, we use tight binding formalism as a framework to understand the
Rashba effects in lead chalcogenide monolayers. Numerical calculations
show that the relevant bands are composed almost exclusively of $s$
and $p$ orbitals of the constituent atoms, with $d$ orbitals appearing
in lower-energy valence bands, allowing us to neglect them (see
Supplemental Material). This means that each atom introduces four (one
$s$ and three $p$) orbitals.  While it is convenient to use $p_x$ and
$p_y$ orbitals to write down the hopping elements, since we are
including SOI in our model, it is helpful to go to a basis which is
more natural for the angular momentum operators. We transform the
basis as follows: $|1,1\rangle = (-|p_x\rangle+i|p_y\rangle)/\sqrt{2}$
and $|1,-1\rangle = (|p_x\rangle+i|p_y\rangle)/\sqrt{2}$.  The new
basis then for each $4\times 4$ block is $|0,0\rangle$, $|1,1\rangle$,
$|1,-1\rangle$, and $|1,0\rangle$, where the first number represents
the orbital momentum quantum number and the second one is the
projection along the $z$ direction.  Details of the Hamiltonian
construction can be found in the supporting information.

\begin{figure}
\includegraphics[width=8cm]{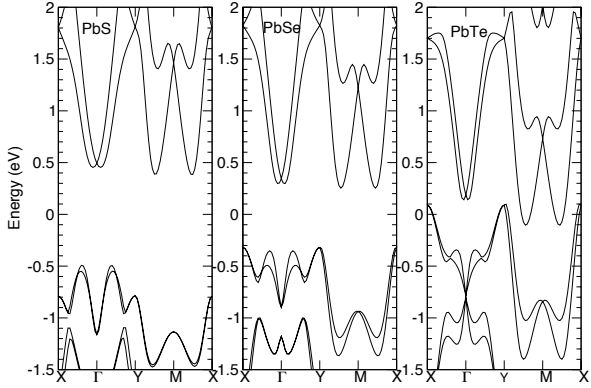}
\caption{Band structures of buckled PbS, PbSe and PbTe with spin-orbit
  interaction included. Fermi energy is set to be zero. All buckled
  lead chalcogenides have large Rashba splitting in the conduction
  band. In the highest valence bands, however, the Rashba splitting is
  smaller for compounds containing lighter chalcogen species. }
\label{fig:bandsCompare}
\end{figure}

To include the SOI, we use the standard form describing the spin-orbit
coupling arising from the interaction with the nucleus
$H_\mathrm{SOI} = T_\mathrm{X}\left(\frac{L_+\otimes s_-+L_-\otimes s_+}{2}+L_z\otimes s_z\right)$,
where X is either Pb or S. The last term modifies the diagonal
elements of the self-energy for $|1,\pm1\rangle$ by adding
(subtracting) $T_\mathrm{X}/2$ if $L_z$ and $s_z$ point in the same
(opposite) direction. The first term couples
$|1,1\rangle\otimes |\downarrow\rangle$ with
$|1,0\rangle\otimes|\uparrow\rangle$ and
$|1,-1\rangle\otimes |\uparrow\rangle$ with
$|1,0\rangle\otimes|\downarrow\rangle$ with the coupling strength
$T_\mathrm{X}/\sqrt{2}$.

The first high-symmetry point that we examine is the $M$-point,
located at $(\pi/2, \pi/2)$ in the Brillouin zone.  At the $M$-point
the full Hamiltonian $H$ can be decomposed into several blocks, and the
Hamiltonian describing the two lowest conduction bands (C1, C2) and the third
valence band (V3) is given by 
\begin{align}
H_b& =\begin{pmatrix}
\varepsilon_p^\mathrm{S}+\frac{T_\mathrm{S}}{2}&\mp4i\alpha^2\Delta&0
\\
\pm4i\alpha^2\Delta&\varepsilon_p^\mathrm{Pb}-\frac{T_\mathrm{Pb}}{2}&\frac{T_\mathrm{Pb}}{\sqrt2}
\\
0&\frac{T_\mathrm{Pb}}{\sqrt2}&\varepsilon_p^\mathrm{Pb}
\end{pmatrix}\,. \label{eq:H_b}
\end{align}
At the $M$-point, the 
degenerate wave functions (labeled as 1 and 2) describing the lowest conduction band C1 are given by
\begin{align}
|\Psi_1\rangle &= 
iA|1,1\rangle\otimes|\uparrow\rangle_\textrm{S}\
+B|1,-1\rangle\otimes|\uparrow\rangle_\textrm{Pb}
+C|1,0\rangle\otimes|\downarrow\rangle_\textrm{Pb}
\nonumber
\\
|\Psi_2\rangle &=-iA |1,-1\rangle\otimes|\downarrow\rangle_\textrm{S}
+B|1,1\rangle\otimes|\downarrow\rangle_\textrm{Pb}
+C|1,0\rangle\otimes|\uparrow\rangle_\textrm{Pb}\,,
\label{eq:orbitals}
\end{align}
where $A$, $B$ and $C$ are real numbers. The other block $H_a$ describing
the highest valence (V1) band has a very similar form to
Eq.~\ref{eq:H_b}, but where Pb and
S are interchanged. 

The degeneracy breaking term $\gamma$ is given by    
\begin{equation}
\gamma=\langle\Psi_1|H|\Psi_2\rangle = 2i\sin\left(2\theta\right)\Delta ke^{i\phi}AC\,,
\label{eq:dirac}
\end{equation}
where $\theta$ is the structure buckling angle, where
$\Delta = V_{pp\sigma}-V_{pp\pi}$ ($V$ is the hopping parameter
between S and Pb atom, see Supplemental Material), and
$ke^{i\phi}=k_x+ik_y$. This leads to a linear dispersion for small
$k$, as expected. Defining
$\lambda\equiv 2\sin\left(2\theta\right)\Delta AC$, we can write the
effective Hamiltonian describing the lowest conduction band as
\begin{equation}
H_{\rm eff}=\lambda[\vec{k}\times\vec{\sigma}]\cdot \hat{z},
\end{equation}
\label{Heff}
where $\vec{\sigma}=(\sigma_x, \sigma_y, \sigma_z)$, which is the
Rashba Hamiltonian. The eigenstates are
$|\psi_{\rm I,II}\rangle=|\Psi_1\rangle\pm ie^{-i\phi}\frac{|\lambda
  |}{\lambda}|\Psi_2\rangle$.

\begin{figure}
\includegraphics[width=8cm]{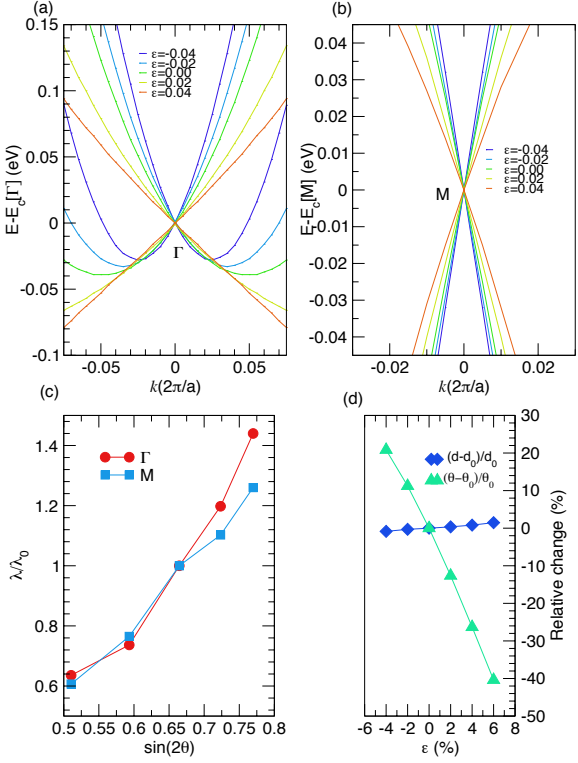}
\caption{Evolution of band structure around $\Gamma$ (a) and $M$ (b)
  with application of biaxial strains.  Energy is subtracted by
  energy at $\Gamma$ ($M$) for comparative purposes. (c) $\lambda$
  scaled by its unstrained value $\lambda_0$ as a function of
  $\sin 2\theta$. $\lambda$ increases with increasing buckling angle,
  which is consistent with tight-binding analysis. (d) Relative
  changes in buckling angle $\theta$ and bond distance $d$ as a
  function of biaxial strain $\epsilon$.}
\label{fig:Er-strains}
\end{figure}

It is clear from Eq.~\ref{eq:H_b} that the SOI mixes the $p_z$ orbital
with other in-plane orbitals of atoms with same species; however SOI
by itself does not lift the degeneracy because SOI is independent of
$k$. For instance, the band structure of {\em planar} PbS obtained by
DFT, including the SOI, does not show spin-splitting
[Fig.~\ref{fig:fig1} (b)]. The inversion symmetry breaking term
originated from the buckling couples the $p_z$ of Pb and the in-plane
$p$ orbitals of S atoms; this term results in the spin-splitting with
Rashba-like dispersion (see Eq.~\ref{eq:dirac}). Taking
$T_{\rm Pb}\gg T_{\rm S}$ and solving the Hamiltonian $H_b$
perturbatively, one can show that, to the first leading order,
$AC\sim T_{\rm Pb}$. These two consequences are consistent with our
DFT results: spin-splitting occurs when both SOI and $\theta$ are not
zero.

While the same arguments hold for $H_a$, which describes the valence
band, we do not observe a substantial SOI-induced splitting in PbS
(see Fig.~\ref{fig:fig1}(c)). This is because the sulphur atom has of
a much smaller atomic SOI than the Pb atom, leading to a weaker mixing
of orbitals, suppressing the $AC$ term in the equation above. As shown
in fig.~\ref{fig:bandsCompare}, PbTe and PbSe, however, have large
spin-splitting in both the conduction and valence bands because Te and
Se are relatively much heavier than S (stronger SOI)~\footnote{The
  optimized geometrical parameters of buckled PbTe and PbSe monolayers
  are tabulated in the supporting information}.

Similarly to the $M$ point, one can perform a low-$k$ expansion around
the $\Gamma$ point for the Hamiltonian matrix (see Supplemental
Material).  Because there are more non-vanishing coupling terms at
the center of the Brillouin zone, the Hamiltonian does not reduce as
well to smaller independent blocks as it does at the $M$
point. Nevertheless, it is possible to show that in buckled
structures, there is a linear term breaking the degeneracy of the
conduction band.

We have found the relevant parameters to tune the band splitting from
the TB-formulation. Clearly the hopping parameters depend on both the
bond distance and the buckling angle. Since these two quantities are
often strain dependent, it is natural to ask whether it is possible to
tune the hopping parameters using strain. Our DFT simulations showed
that under biaxial strains the bond distance changes by only 1\% while
buckling angle changes by roughly 30\% at a biaxial strain of 6\%
(shown in Fig.~\ref{fig:Er-strains}).

We obtain $\lambda$ by taking the derivative of energy dispersion
$\lambda=\frac{\partial E}{\partial k}$ at $\Gamma$ and $M$. As shown
in Fig.~\ref{fig:Er-strains} (c), $\lambda$ increases with increasing
$\theta$, consistent with our TB formulation (see
Eq.~\ref{eq:dirac}). Note that $\lambda$ is not linear with
$\sin 2\theta$ because $A$ and $C$ also depend on $\theta$. Our DFT
results show that, relative to its unstrained value $\lambda_0$,
$\lambda$ can increase by more than 20\% when compressed by 4\% or
decrease by 20\% when stretched by 4\%. The apparent variations of
$\lambda$ show that PbS is a tunable spin-splitting material.

\emph{Spin and Orbital Texture--} 
\begin{figure*}
\includegraphics[width=16cm]{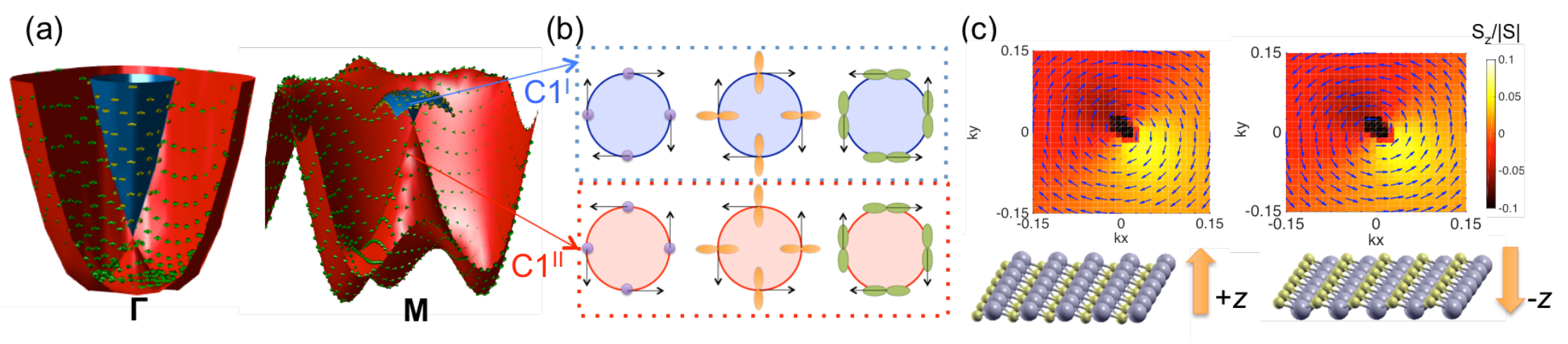}
\caption{ (a) Band plots of the first (C1$^{\rm II}$) and second
  lowest (C1$^{\rm I}$) conduction band near the $\Gamma$ and $M$
  point. Clockwise (counter clockwise) spin textures are represented by
  the yellow (green) arrows. Near the band crossing (inner Dirac
  cone), the upper and lower band have opposing helical spin texture
  similar to the Rashba spin texture. (b) Corresponding schematic orbital
  spin texture of Pb atom at $M$ point. The radial $p_r$ and
  tangential $p_t$ have opposite spin orientation, and they cancel each
  other. Spin helicity is flipped after passing through the Dirac
  point while the orbital compositions are still the same.  (c) Two
  dimensional plot of spin polarizations near $M$. The color plot
  shows the projection of spin along $z$ direction. Clearly, the
  out-of-plane spin components are small. The direction of spin
  polarizations is reversed when the buckling direction is
  reversed. }
\label{fig:spinTexture}
\end{figure*}
Lastly, we investigate the orbital texture of PbS as it has been shown
that TIs and hexagonal 3D Rashba materials have orbital switching at
the Dirac point~\cite{cao-NatPhys-9-499-2013,
  zhang-PRL-111-066801-2013, liu-PRB-94-125207-2016,
  bawden-scieAdv-1-e1500495-2015}. To our best knowledge, such
analysis has not been done for monolayers with square symmetry.  We
can do such analysis by transforming our basis to radial $p_r$ and
tangential $p_t$ orbitals:
$|1, 1\rangle=-e^{i\phi}|p_r\rangle+i|p_t\rangle$,
$|1, -1\rangle=e^{-i\phi}|p_r\rangle-i|p_t\rangle$ and
$|1, 0\rangle=|p_z\rangle$. In this basis the upper and bottom
wavefunctions can be written as 
\begin{align}
|\psi_{\rm I,II}\rangle =&C|p_z\rangle\otimes|\pm\rangle_{\rm Pb}
\mp i\frac{B}{\sqrt2}|p_r\rangle\otimes|\pm\rangle_{\rm Pb}\\
\nonumber
&\pm \frac{B}{\sqrt2}|p_t\rangle\otimes|\mp\rangle_{\rm Pb}
\mp i\frac{A}{\sqrt2}e^{-2i\phi}|p_r\rangle\otimes|\mp_3\rangle_{\rm S}\\
\nonumber
&\pm \frac{A}{\sqrt2}e^{-2i\phi}|p_t\rangle\otimes|\pm_3\rangle_{\rm S}
\label{eq:spinorbitals}
\end{align}
where $|+_n\rangle=\frac{1}{\sqrt2}\begin{pmatrix}
  \frac{|\lambda|}{\lambda} ie^{-i n\phi}\\
  1\end{pmatrix}$
is clockwise in-plane spin and
$|-_n\rangle=\frac{1}{\sqrt2}\begin{pmatrix}
  -\frac{|\lambda|}{\lambda} ie^{-i n\phi}\\
  1\end{pmatrix}$ is counter clockwise in-plane spin.

In the upper band of PbS the radial component of the Pb atom couples to
the clockwise spin while the tangential component couples to the counter
clockwise spin, as shown schematically in Fig.~\ref{fig:spinTexture}. As
it passes through the band crossing point (Dirac point) right at the
$M$ point, where the upper band and lower band meet, the tangential
component now couples to the counter clockwise spin. This switching is
similar to what have been observed in
TIs~\cite{cao-NatPhys-9-499-2013, zhang-PRL-111-066801-2013} and
hexagonal bulk Rashba materials~\cite{bawden-scieAdv-1-e1500495-2015,
  liu-PRB-94-125207-2016}. The difference is that the radial and
tangential components contribute equally and cancel out, and thus
the net in-plane spin texture comes from the
$p_z$ orbital only. This suggests that the orbital texture is not
always polarized and thus the orbital polarization depends on the
crystal symmetry of the material.

From the TB results, we found that the direction of the spin is given by
$\langle\psi_{\rm I, II}|\hat{\sigma}|\psi_{\rm I,
  II}\rangle=\pm\frac{|\lambda|}{\lambda}(\sin\phi, -\cos\phi, 0)$.
We can see that the spin texture is helical and depends on the
direction of the buckling (inversion symmetry breaking term
$\lambda$).  Notice that the PbS-$b$ has a degenerate structure as the
polarization vector in $z$-direction define as
$d_z=z({\rm Pb})-z({\rm S})$ can be positive or negative (sign of
$\lambda$), as shown in Fig.~\ref{fig:spinTexture} (c). Thus, the
in-plane spin texture can be reversed when $\lambda$ is negative (PbS
buckled in the opposite direction). This is confirmed by our DFT
results shown in Fig.~\ref{fig:spinTexture}(c). While the coefficients
$A$, $B$, and $C$ in Eq.~\ref{eq:orbitals} are material dependent, the
orbital texture is independent of the direction of the buckling. These
findings are in agreement with the very recent work on hexagonal 3D
Rashba BiTeI~\cite{liu-PRB-94-125207-2016}.

\emph{Conclusion--} In summary, using first-principles calculations
based on density functional theory, we have found a new class of 2D
materials (lead chalcogenides) possessing a tunable giant Rashba
splitting with a characteristic orbital and spin texture in an energy
range close to the bandgap edge. Based on our tight-binding analysis,
we found that the atomic composition and buckling angle are the two
key parameters controlling the Rashba effects. First, the atomic
composition plays important role as the SOI is the parameter that
mixes the in-plane and out-of-plane orbitals. With the recent success
in creating janus (polar) transition metal dichalcogenide
monolayers~\cite{lu-NatNano-2017} and few-layer SnTe and
PbTe~\cite{chang-science-353-274-2016} via atomic layer deposition
techniques, the growth of buckled polar materials like PbS, PbSe, and
PbTe should also be achievable using existing technology.

In buckled PbS monolayers, the Rashba coefficient depends on the
degree of buckling, and the orientation of the helical in-plane spin
depends on the direction of the buckling.  As we have shown in DFT
simulations, this buckling can be controlled through application of
moderate strains of $\leq10\%$, which are achievable in the current 2D
experiments~\cite{lloyd-NL-16-5836-2016, perez-NL-14-4107-2014}. A
similar system showing such tunability is
LaOBiS$_2$~\cite{qihang-NL-13-5264-2013}. In addition to mechanical
strains, the electric polarization (direction of buckling) can be
switched as PbS is ferroelectric. And thus PbS spin texture can be
switched in a non-volatile way which is similar to recently found
ferroelectric Rashba semiconductors (FERSC)
GeTe~\cite{sante-AdvMat-25-1521-2013, sante-AdvMat-25-1521-2013}.
Further, we also found orbital-spin texture switching in buckled
PbS. Our results suggest that the orbital-spin switching at the Dirac
point is not exclusive to TIs and the orbital texture is not always
polarized, as it depends on the crystal symmetry of the material. Our
unifying framework based on tight binding provides design principles
and orbital-spin texture manipulations which will be important for
development of new devices.

\begin{acknowledgements}
  P.Z.H., H.S.P. and D.K.C. acknowledge the support of the Physics and
  Mechanical Engineering Department at Boston University. P.Z.H.  is
  grateful for the hospitality of the NUS Centre for Advanced 2D
  Materials and Graphene Research Centre where this work was
  initiated.  D.K.C. acknowledges the hospitality of the Aspen Center
  for Physics, which is supported by the U.S. National Science Foundation grant
  PHY-1607611. A.C. acknowledges support by the National Research
  Foundation, Prime Minister Office, Singapore, under its Medium Sized
  Centre Programme and CRP award ``Novel 2D materials with tailored
  properties: beyond graphene'' (Grant No.  R-144-000-295-281).
\end{acknowledgements}

\bibliography{rashba_RPRB}

\pagebreak
\widetext
\begin{center}
\textbf{\large Supplemental Materials}
\end{center}

\section{Tight binding}
\begin{figure}
\includegraphics[width=14cm]{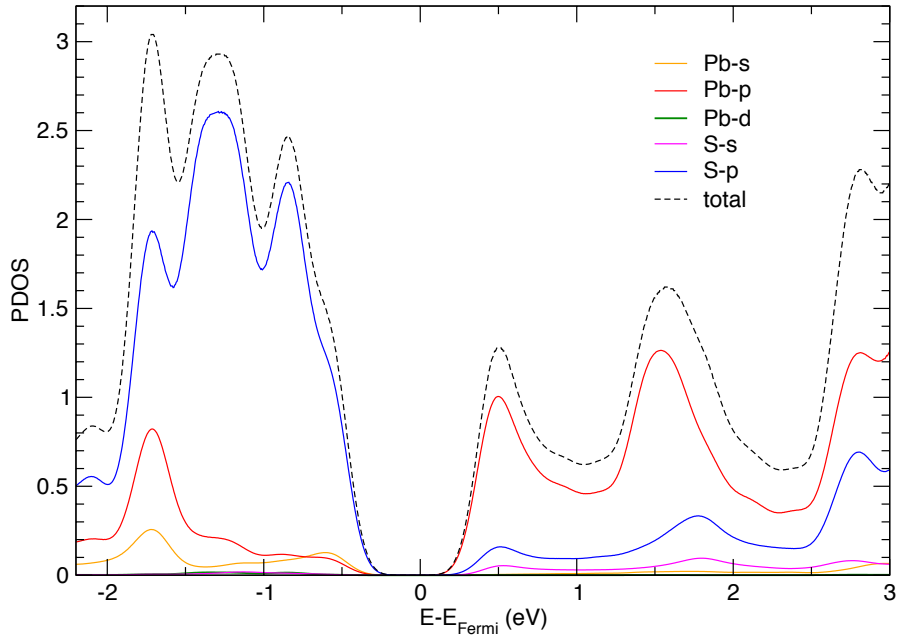}
\caption{Projected density of states (PDOS) of Pb and S atoms. Note
  that the Pb-d contribution is too small to be seen on this figure}
\label{fig:SI_fig1}
\end{figure}
The \emph{ab initio} calculations reveal a
fairly complicated band structure with multiple local minima in the
conduction band and a fairly flat valence band. In order to better
understand the origin of the band structure, we turn to the
tight-binding model.

Numerical calculations show that the relevant bands are composed
almost exclusively of $s$ and $p$ orbitals of the constituent atoms,
with $d$ appearing in lower-energy valence bands, allowing
us to neglect them (see fig.~\ref{fig:SI_fig1}). The lowest conduction
band consists mostly the $p$ orbitals of the Pb atom while the highest
valence band consists mostly the $p$ orbitals of the S atom.

This means that each atom introduces four (one $s$
and three $p$) orbitals. Accounting for the spin and two atomic
species leads to a sixteen-component basis. Setting the bond length
between the neighboring Pb and S atoms to $d$, the
vectors connecting a Pb atom to its four neighbors
are
\begin{equation}
\mathbf{d}_n=d
\Bigg(\cos\theta\cos \vartheta_n,\cos\theta \sin\vartheta_n,-\sin\theta\Bigg),
\end{equation}
where $\theta$ determines the buckling angle (with $\theta=0$
corresponding to a flat lattice) and $\vartheta_n$ is the azimuthal
angle, $\vartheta_n=\frac{\pi}{2}(n+\frac{1}{2})$ with $n=0, 1, 2, 3$.
We denote $\alpha =\cos\theta/\sqrt 2$, $\beta = \sin\theta$.  The
size of the unit cell is $2\alpha d\times 2\alpha d$ so that the first
Brillouin zone extends for
$-\frac{\pi}{2\alpha d}\leq q_x,q_y\leq\frac{\pi}{2\alpha d}$.

Since PbS is formed by two shifted square lattices, we first consider
the Hamiltonian describing a single atomic species. In the absence of
SOI, the basis is $|s\rangle$, $|p_x\rangle$, $|p_y\rangle$, and
$|p_z\rangle$. Due to the planar structure of the sublattice, $p_z$
does not couple to other orbitals. The scaled momentum
$k_{x/y} = q_x \alpha d$ ranges between $-\pi/2$ and $\pi/2$ and
$\bar {E}$ is a diagonal matrix with on-site orbital energies
$(\epsilon_s, \epsilon_p, \epsilon_p, \epsilon_p)$ on the diagonal. 

The next step is to introduce coupling between the two sublattices. As
the hopping between Pb and S atoms does not flip spins, only same-spin
blocks are coupled. This leads to a straightforward, if bulky, hopping
block, see Eq.~\eqref{eqn:K}. There are five parameters here:
$V_{ss\sigma}$, $V_{sp\sigma}^{(1)}$, $V_{sp\sigma}^{(2)}$,
$V_{pp\sigma}$, and $V_{pp\pi}$. They describe (in order): hopping
between $s$ orbitals of the different atomic species, hopping between
$s$ orbital of Pb and $p$ orbital of S, hopping between $s$ orbital of
S and $p$ orbital of Pb, hopping between $\sigma$ oriented $p$
orbitals of the two species, and hopping between $\pi$ oriented $p$
orbitals of the two species.

%
\begin{align}
K &= 4\cos\left(k_x\right)\cos\left(k_y\right)\begin{pmatrix}
V_{ss\sigma}&0&0&-\beta V_{sp\sigma}^{(1)}
\\
0&\left[\alpha^2\Delta+V_{pp\pi}\right]&0&0
\\
0&0&\left[\alpha^2\Delta+V_{pp\pi}\right]&0
\\
\beta V_{sp\sigma}^{(2)}&0&0&\left[\beta^2\Delta+V_{pp\pi}\right]
\end{pmatrix}
\nonumber
\\
 +&4\sin\left(k_x\right)\sin\left(k_y\right)\begin{pmatrix}
0&0&0&0
\\
0&0&-\alpha^2\Delta&0
\\
0&-\alpha^2\Delta&0&0
\\
0&0&0&0
\end{pmatrix}
\nonumber
\\
+& 4\cos\left(k_y\right)\sin\left(k_x\right)\begin{pmatrix}
0&i\alpha V_{sp\sigma}^{(1)}&&0
\\
-i\alpha V_{sp\sigma}^{(2)}&0&0&-i\alpha\beta\Delta
\\
0&0& 0&0
\\
0&-i\alpha\beta\Delta&0&0
\end{pmatrix}
\nonumber
\\
+&  4\cos\left(k_x\right)\sin\left(k_y\right)\begin{pmatrix}
0&0&i\alpha V_{sp\sigma}^{(1)}&0
\\
0&0&0&0
\\
-i\alpha V_{sp\sigma}^{(2)}&0& 0&-i\alpha\beta\Delta
\\
0&0&-i\alpha\beta\Delta&0
\end{pmatrix}\,,
\label{eqn:K}
\end{align}

While it is convenient to use $p_x$ and $p_y$ orbitals to write down
the hopping elements, since we are including SOI in our model, it is
helpful to go to a basis which is more natural for the angular
momentum operators. We transform the basis as follows:
$|1,1\rangle = (-|p_x\rangle+i|p_y\rangle)/\sqrt{2}$ and
$|1,-1\rangle = (|p_x\rangle+i|p_y\rangle)/\sqrt{2}$.
The new basis then for each $4\times 4$ block is $|0,0\rangle$,
$|1,1\rangle$, $|1,-1\rangle$, and $|1,0\rangle$, where the first
number represents the orbital momentum quantum number and the second
one is the projection along the $z$ direction.

To include the SOI, we use the standard form describing the spin-orbit
interaction arising from the interaction with the nucleus:
\begin{equation}
H_\mathrm{SOI} = T_\mathrm{X}\left(\frac{L_+\otimes s_-+L_-\otimes s_+}{2}+L_z\otimes s_z\right)\,,
\label{eqn:SOI}
\end{equation}
where X is either Pb or S. The last term modifies the diagonal
elements of the self-energy for $|1,\pm1\rangle$ by adding
(subtracting) $T_\mathrm{X}/2$ if $L_z$ and $s_z$ point in the same
(opposite) direction. The first tem couples
$|1,1\rangle\otimes |\downarrow\rangle$ with
$|1,0\rangle\otimes|\uparrow\rangle$ and
$|1,-1\rangle\otimes |\uparrow\rangle$ with
$|1,0\rangle\otimes|\downarrow\rangle$ with the coupling strength
$T_\mathrm{X}/\sqrt{2}$.

\subsection{M Point}
\begin{figure}
\includegraphics[width=14cm]{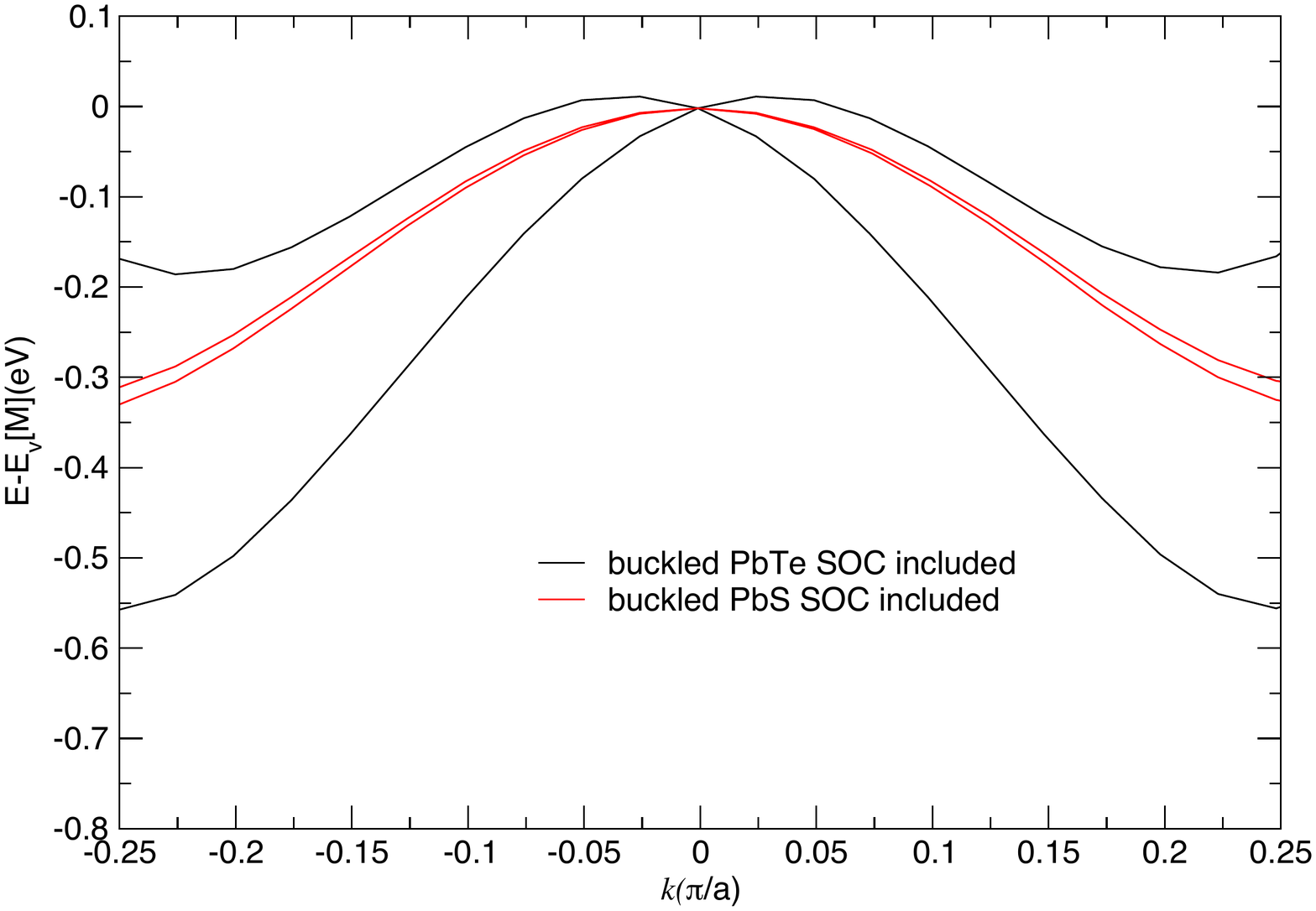}
\caption{Valence band around $M$ for buckled PbS and PbTe with SOI
  included. The band is shifted for comparative purpose. }
\label{fig:SI_fig2}
\end{figure}

The first high-symmetry point that we turn to is the $M$-point, located at  $(\pi/2, \pi/2)$ in the Brillouin zone. At the M-point, the full Hamiltonian decomposes into several blocks:
%
\begin{align}
H_1 &= \begin{pmatrix}
\varepsilon_s^\mathrm{Pb}
\end{pmatrix}\,,\quad H_2 = \begin{pmatrix}
\varepsilon_s^\mathrm{S} 
\end{pmatrix}\,,
\\
H_a&=\begin{pmatrix}
\varepsilon_p^\mathrm{Pb}+\frac{T_\mathrm{Pb}}{2}&0&\pm4i\alpha^2\Delta
\\
0&\varepsilon_p^\mathrm{S}&\frac{T_\mathrm{S}}{\sqrt2}
\\
\mp4i\alpha^2\Delta&\frac{T_\mathrm{S}}{\sqrt2}&\varepsilon_p^\mathrm{S}-\frac{T_\mathrm{S}}{2}
\end{pmatrix}\,,
\\
H_b& =\begin{pmatrix}
\varepsilon_p^\mathrm{S}+\frac{T_\mathrm{S}}{2}&\mp4i\alpha^2\Delta&0
\\
\pm4i\alpha^2\Delta&\varepsilon_p^\mathrm{Pb}-\frac{T_\mathrm{Pb}}{2}&\frac{T_\mathrm{Pb}}{\sqrt2}
\\
0&\frac{T_\mathrm{Pb}}{\sqrt2}&\varepsilon_p^\mathrm{Pb}
\end{pmatrix}\,,
\end{align}
%
where $S$ and $P$ denote the hopping in sulphur and lead sublattices
and $T_\mathrm{S}$ ($T_\mathrm{Pb}$) is the SOI coefficient in sulphur
(lead). The first two blocks are doubly degenerate, describing spin-up
and spin-down states composed entirely of $s$ orbitals on both atomic
species. The degenerate $H_a$ acts over the sets of wavefunctions:
($|1,1\rangle\otimes|\uparrow\rangle_\mathrm{Pb}$,
$|1,0\rangle\otimes|\downarrow\rangle_\mathrm{S}$, and
$|1,-1\rangle\otimes|\uparrow\rangle_\mathrm{S}$) and
($|1,-1\rangle\otimes|\downarrow\rangle_\mathrm{Pb}$,
$|1,0\rangle\otimes|\uparrow\rangle_\mathrm{S}$, and
$|1,1\rangle\otimes|\downarrow\rangle_\mathrm{S}$). $H_b$, on the
other hand, mixes the following sets of three states:
($|1,-1\rangle\otimes|\downarrow\rangle_\mathrm{S}$,
$|1,1\rangle\otimes|\downarrow\rangle_\mathrm{Pb}$ and
$|1,0\rangle\otimes|\uparrow\rangle_\mathrm{Pb}$) and
($|1,1\rangle\otimes|\uparrow\rangle_\mathrm{S}$,
$|1,-1\rangle\otimes|\uparrow\rangle_\mathrm{Pb}$ and
$|1,0\rangle\otimes|\downarrow\rangle_\mathrm{Pb}$).

$H_a$ results in three doubly-degenerate bands: V2, V1, and C3; $H_b$,
in turn, gives rise to V3, C1, and C2; where V and C represent valence
and conduction bands, respectively. In a flat lattice, the valence
(conduction) band is composed exclusively of $p_z$ orbitals of sulphur
(lead) atoms. As can be seen in Fig. 1 (b) and (c) of the manuscript,
without SOI, both the valence and the conduction bands are concave
down at the $M$ point. In fact, the variation in the band energy along
the X-M-Y line arises only due to the hopping between $p_z$ orbitals
of same-species atoms. Naturally, all the bands are degenerate in this
case. Introducing buckling switches on the SOI and lifts the
degeneracy in the following manner.

The degenerate wave functions (labeled as 1 and 2) describing C1 exactly at M are given by
\begin{align}
|\Psi_1\rangle &= 
iA|1,1\rangle\otimes|\uparrow\rangle_\textrm{S}\
+B|1,-1\rangle\otimes|\uparrow\rangle_\textrm{Pb}
+C|1,0\rangle\otimes|\downarrow\rangle_\textrm{Pb}
\\
|\Psi_2\rangle &=-iA |1,-1\rangle\otimes|\downarrow\rangle_\textrm{S}
+B|1,1\rangle\otimes|\downarrow\rangle_\textrm{Pb}
+C|1,0\rangle\otimes|\uparrow\rangle_\textrm{Pb}\,,
\end{align}
where $A$, $B$ and $C$ are real numbers.

The degeneracy breaking term is given by
\begin{equation}
\langle\Psi_1|H|\Psi_2\rangle = 2i\sin\left(2\theta\right)\Delta ke^{i\phi}AC\,,
\label{eq:dirac_SI}
\end{equation}
leading to a Dirac dispersion for small $k$, as expected. Defining
$\lambda\equiv 2\sin\left(2\theta\right)\Delta AC$, we can write the
effective Hamiltonian describing C1 as
\begin{equation}
H_{\rm eff}=\lambda[\vec{k}\times\vec{\sigma}]\cdot \hat{z}
\end{equation}
\label{Heff_SI}
which is the Rashba Hamiltonian. The eigenstates are
$|\psi_{\rm I,II}\rangle=|\Psi_1\rangle\pm ie^{-i\phi}\frac{|\lambda
  |}{\lambda}|\Psi_2\rangle$.

As we discussed in the manuscript, $H_b$ describes the lowest
conduction band C1 while $H_a$ describes the highest valence band
V1. As $AC\sim T_{\rm S}$ in $H_a$, the splitting in the valence
band of PbS is weak. PbTe, however, has large splittings in both
conduction and valence band because Te is relatively much heavier than
S atom (stronger SOI). This is again confirmed by DFT results (see
fig.~\ref{fig:SI_fig2}).

\section{Bistability}
\begin{figure}
\includegraphics[width=14cm]{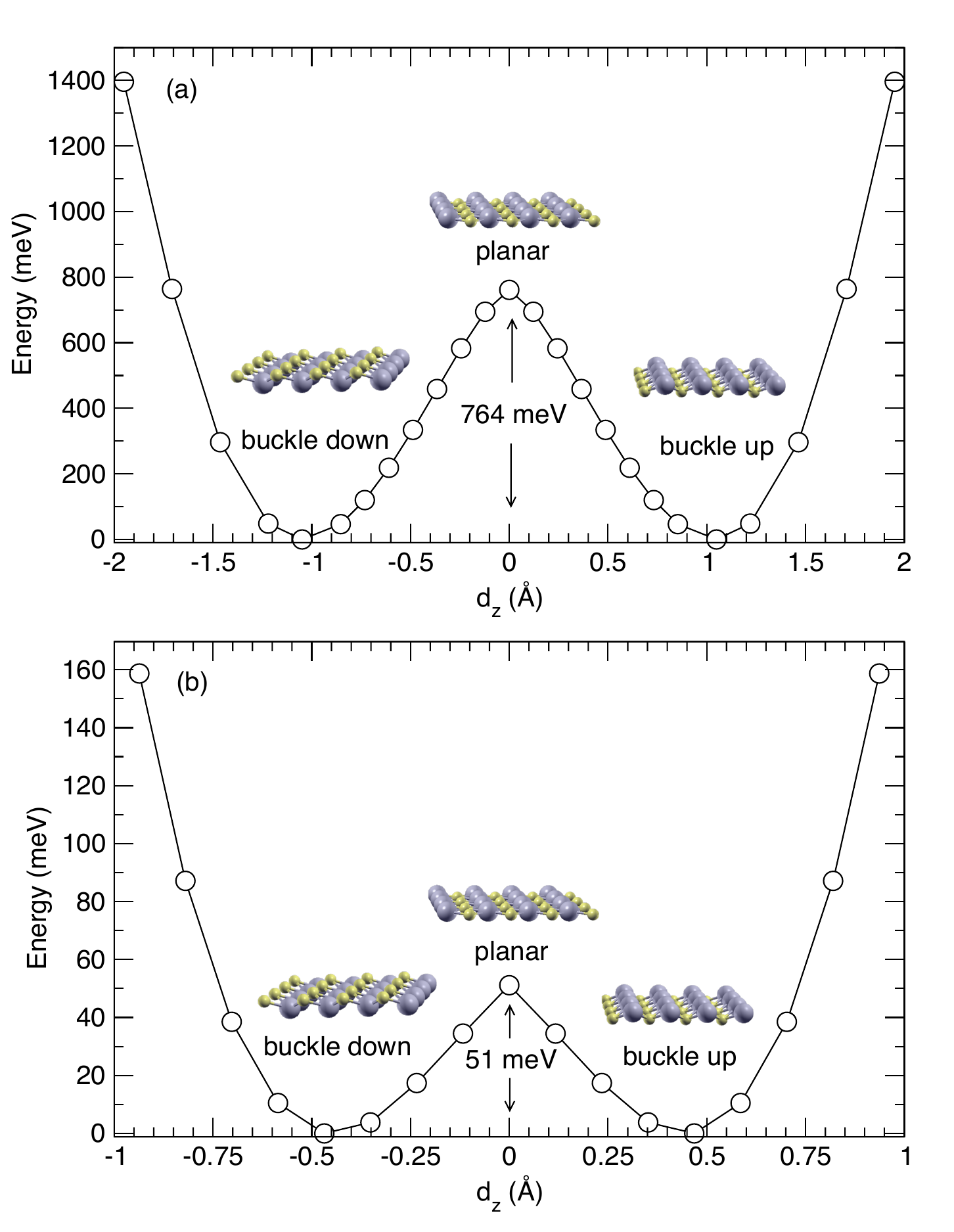}
\caption{Potential energy surface of PbS as a function of distance
  $d_z$ with (a) fixed buckled lattice parameters and (b) fixed planar
  lattice parameters.}
\label{fig:SI_fig3}
\end{figure}

We created a vacuum region of 20\AA~ perpendicular to the plane to
model the monolayer. The in-plane lattice parameters of PbS monolayers
are optimized using density functional theory (DFT). We found that PbS
has a buckled structure, which is a minimum of the energy surface,
whereas the planar structure is a saddle point of the energy
surface. The {\it optimized} planar structure of PbS has a higher
enthalpy of 120~meV compared to that of {\it optimized} buckled
structure. The lattice constant $a$ and buckling angle $\theta$ for
buckled (planar) structure are 3.74\AA~(4.07\AA) and 21.6$^{\circ}$
(0$^{\circ}$), respectively.

It is known that potential barriers in bistable materials, such as
ferroelectrics, are strain
dependent~\cite{wang-2DMat-4-015042-2017}. We follow procedure by
Ref.~\cite{wang-2DMat-4-015042-2017} to obtain the potential energy
surface. We vary $d_z=z({\rm Pb})-z({\rm S})$ with the fixed optimized
buckled lattice parameters. In this case, the energy of planar
structure is higher than the optimized planar structure, as we do not
relax the lattice constants (see fig.~\ref{fig:SI_fig3} (a)). In
addition, we also do the same procedure but using the planar structure
as the initial configuration. In the second case, the lattice
parameters are fixed at the optimized planar lattice parameters (see
fig.~\ref{fig:SI_fig3} (b)). In either case, we find that the planar
structure is a saddle point.

We used the modern theory of
polarization to calculate electric polarization\cite{vanderbilt-PRB-47-1651-1993}
\begin{equation} 
\vec{\mathcal{P}}=\frac{1}{\Omega}\sum_{\tau}q^{\rm{ion}}_{\tau}{\bf R}_{\tau}-\frac{2i\rm{e}}{(2\pi)^3}\sum_{n}^{\rm{occ}}\int_{BZ}d^3{\bf k}e^{-i\vec{k}\cdot{\bf R}}\Big\langle u_{n{\bf k}}\Big|\frac{\partial u_{n{\bf k}}}{\partial {\bf k}}\Big\rangle, 
\label{eq:pol}
\end{equation}
where $q_\tau$ is the ionic charge plus the core electrons,
${\bf R}_{\tau}$ is the position of ions, $\Omega$ is the unit cell
volume, $\rm{e}$ is the elementary charge, $n$ is the valence band
index, ${\bf k}$ is the wave vector, and $u_{n{\bf k}}$ is the
electronic wave function. The first term is the contribution from core
electrons and ions, and the second term is the electronic contribution
defined as adiabatic flow of current which can be calculated from
Berry connection~\cite{vanderbilt-PRB-47-1651-1993}. To calculate the
electric polarization of monolayer PbS, we estimate the thickness as
twice the distance between S and Pb atom which is roughly half of the
lattice constant of bulk PbS.  Similar approximations have also done
in other several works~\cite{hanakata-PRB-94-035304-2016,
  fei-PRL-117-097601-2016, wang-2DMat-4-015042-2017}.The calculated
spontaneous polarizations of path I (optimized buckled lattice
parameters) and path II (optimized planar lattice parameters) are
0.2~C/m$^2$ and 0.1~C/m$^{2}$, respectively.

\section{Lead Chalcogenides Optimized Structure}
Here we tabulate the optimized geometrical parameters of buckled PbX (X=S, Se, and Te) monolayers. 
\begin{table}[h]
\small
  \caption{Lattice constant $a$, buckling angle $\theta$, buckling height $d_z$, and nearest-neighbor bond distance $d$. }
\centering
\begin{tabular*}{0.5\textwidth}{@{\extracolsep{\fill}}lllll}
\hline 
& $a${\rm ~\AA} & $\theta(^\circ)$& $d_z${\rm ~\AA} & $d${\rm ~\AA} \\
\hline 
PbS   &   3.74 &   21.6 &  1.97  & 2.84  \\
PbSe    & 3.82 &  24.4    & 2.31    & 2.96 \\
PbTe    & 4.01   & 26.3     & 2.65 & 3.16\\
\hline
\end{tabular*}
\label{table:comparison}
\end{table}

\end{document}